\documentclass[12pt]{iopart}
\usepackage{iopams}

\usepackage{graphicx}

\renewcommand{\Tr}[1]{\mbox{Tr}\,#1}
\renewcommand{\tr}[1]{\mbox{tr}\,#1}
\newcommand{\ket}[1]{| #1\rangle}

\newcommand{\proj}[2]{| #1\rangle\langle #2|}
\newcommand{\ipro}[2]{\langle #1|#2 \rangle}
\newcommand{\expec}[1]{\langle #1 \rangle}
\newcommand{\Eop}{\hat{E}}

\newcommand{\Iop}{\hat{I}}

\newcommand{\Kop}{\hat{K}}

\newcommand{\Mop}{\hat{M}}

\begin{document}
\title{Entanglement amplification via local weak measurements}

\author{Yukihiro Ota}
\address{
Advanced Science Institute, RIKEN, Wako-shi, Saitama, 351-0198, Japan}
\ead{otayuki@riken.jp}

\author{Sahel Ashhab}
\address{
Advanced Science Institute, RIKEN, Wako-shi, Saitama, 351-0198, Japan}
\address{
Physics Department, University of Michigan, 
Ann Arbor, Michigan 48109-1040, USA}

\author{Franco Nori}
\address{
Advanced Science Institute, RIKEN, Wako-shi, Saitama, 351-0198, Japan}
\address{
Physics Department, University of Michigan, 
Ann Arbor, Michigan 48109-1040, USA}

\begin{abstract}
We propose a measurement-based method to produce a maximally-entangled
 state from a partially-entangled pure state. 
Our goal can be thought of as entanglement distillation from a single
 copy of a partially-entangled state.  
The present approach involves local two-outcome weak measurements. 
We show that application of these local weak  
measurements leads to a probabilistic amplification of entanglement. 
In addition, we examine how the probability to find the
 maximally-entangled state is related to the entanglement of the
 input state. 
We also study the application of our method to a mixed initial state. 
We show that the protocol is successful if the separable part of the
 mixed initial state fulfils certain conditions. 
\end{abstract}


\maketitle

\section{Introduction}
Measurements on quantum systems as well as the coherent
manipulation via unitary operations are key ingredients to the
implementation of interesting quantum state-engineering protocols. 
An essential feature of quantum
measurements\,\cite{Davies:1976,Kraus:1983,Peres:1993,DEspagnat:1999} 
is that these significantly affect the static and dynamic properties
of the system. 
This allows phenomena not predicted by classical mechanics, such as 
the quantum Zeno
effect\,\cite{Misra;Surdarshan:1977,Nakazato;Pascazio:1996}.  

Many measurement-based schemes have
been studied including: one-way quantum
computation\,\cite{Raussendorf;Briegel:2003}, 
protection or recovery of a quantum state in a noisy
channel\,\cite{Facchi;Pascazio:2004,Sun;Zubairy:2010,Korotkov;Keane:2010,PazSilva;Lidar:2011},    
preparation of an entangled state\,\cite{Yuasa;Nakazato:2009}, 
and measurement-based quantum
control\,\cite{Branczyk;Bartlett:2007,Ashab;Nori:2010}. 
Among these approaches, the use of weak (or general) measurements has
recently drawn considerable attention.  
General measurements are the generalization of von Neumann
measurements and are associated with a positive-operator valued
measure (POVM)\,\cite{Davies:1976,Kraus:1983,Peres:1993}. 
Their influence on a quantum system is, in general, more moderate than
von Neumann measurements. 
Thus, in some cases, the amount of information extracted using weak  
measurements can be tunable. 
This feature is useful to implement quite interesting quantum protocols, 
such as 
reversing measurements\,\cite{Koashi;Ueda:1999,Kim;Kim:2009} 
and weak-value measurements\,\cite{Iinuma;Hofmann:2011,Kofman;Nori:2011}. 


In this paper, we study a method to prepare a maximally-entangled state
from a pure initial state using weak measurements. 
Our goal can be considered to be entanglement distillation with a single
copy of a partially-entangled state. 
The present approach involves local two-outcome weak measurements. 
We show that the application of local weak
measurements leads to a probabilistic amplification of the entanglement. 
Furthermore, we examine the relationship of our scheme with an
entanglement-concentration
protocol\,\cite{Bennett;Schumacher:1996,Kwiat;Gisin:2001}. 
The present study shows several interesting roles of a weak measurement. 

The paper is organized as follows. 
First, we explain our setting in section \ref{sec:setting}. 
Section \ref{sec:results} is the main part of this paper. 
We propose a method to produce a maximally-entangled state
from a partially-entangled pure state with the application of local weak
measurements. 
The key idea is to tune the measurement strength at each measurement
step. 
Our approach developed in section \ref{sec:results} assumes that the
input state is a pure state.  
In order to consider more realistic situations, we examine the case when
the input state is a mixed state in section \ref{sec:mixed}. 
Then, we find that the method for a pure state still leads to
entanglement amplification if the separable part of the initial mixed
state fulfills certain conditions. 
Furthermore, we discuss the extendability of this approach in section
\ref{sec:discussion}.   
Section \ref{sec:summary} is devoted to a summary of the results.  

\section{Setting}
\label{sec:setting}
\subsection{Linear entropy}
Let us consider a bipartite system composed of a two- and
a $d$-level system ($d \ge 2$). 
The two-level system is called system A, while the $d$-level one
is called system B. 
We will apply some operations only to system A and never touch system B.  
Since system A is a single-qubit, the treatment of control
processes is rather simple.  
We have a pure state $\ket{\psi}$ in this composite system, 
\begin{equation}
 \ket{\psi} 
= 
\alpha \ket{0}\ket{\phi_{0}} + \beta \ket{1}\ket{\phi_{1}},
\quad
\ket{i} \in \mathbb{C}^{2},
\quad
\ket{\phi_{i}}\in \mathbb{C}^{d},
\label{eq:input}
\end{equation}
with $\alpha^{2}+\beta^{2}=1$, $\alpha \ge 0$, $\beta \ge 0$, 
$\ipro{i}{j}=\delta_{ij}$, and 
$\ipro{\phi_{i}}{\phi_{j}}=\delta_{ij}$ ($i,j=0,1$). 
We also note that $\alpha$ and $\beta$ are real parameters. 
Mathematically, we can always find this form of a vector with the
Schmidt decomposition~\cite{Peres:1993}. 
The quantum correlation of $\ket{\psi}$ is characterized by the purity
of the reduced density matrix   
\(
\rho_{\rm A} = \Tr_{\rm B} (\proj{\psi}{\psi})
\), as seen in, e.g., \cite{DEspagnat:1999}. 
We consider the linear entropy  
\begin{equation}
 S(\psi)
=
2[1 - \Tr_{\rm A} (\rho_{\rm A}^{2})]
=
4\alpha^{2}\beta^{2}.
\label{eq:l_entropy}
\end{equation}
When $S(\psi)=0$, we find that either $\alpha=0$ or $\beta=0$. 
Hence, there is no correlation between the two subsystems.  
In contrast, when $S(\psi) > 0$, $\ket{\psi}$ has a
quantum correlation, i.e., entanglement. 
In particular, $S(\psi)$ takes its maximum value $1$ when
$\rho_{\rm A}$ is a maximally-mixed state (i.e., when $\alpha=\beta$).  
This indicates that we have a maximally-entangled state. 
Thus, the linear entropy $S(\psi)$ can be regarded as an entanglement
measure. 

\subsection{Motivation} 
Entanglement is one of the key resources of quantum information processing
such as quantum teleportation, quantum dense coding, quantum key
distribution, and so on\,\cite{Brub;Leuchs:2007}. 
To develop a way to create and protect maximally-entangled states is
quite important because most quantum protocols rely on their existence. 
Thus, an interesting physical issue is that (\ref{eq:input})
is a partially-entangled state (i.e., $\alpha \neq \beta$ and
$\alpha,\beta>0$) and one might desire to amplify its entanglement. 
Typically, one can encounter such an issue when one tries to create a
maximally-entangled state in the presence of experimental
imperfections. 
Our aim is to find an efficient and practical way to transform
$\ket{\psi}$ into a maximally-entangled state. 
The most straightforward way might be to use controlled unitary
operators between the two subsystems. 
Entanglement purification with many
copies of $\ket{\psi}$ is also a candidate to accomplish this goal, as
seen, e.g., in \cite{Brub;Leuchs:2007}. 
However, these methods can be costly, e.g., either requiring properly
controlled gates, or preparing a sufficient number of copies, or
two-body measurements. 
Here we wish to propose a simpler method to achieve the same goal. 
Namely, we want to create a maximally-entangled state
from a partially-entangled state using a simpler experimental setting. 
Thus, we consider the case when one can use only {\em local} operations
and a {\em single} copy of $\ket{\psi}$. 

We now illustrate our task more explicitly. 
Without loss of generality, we assume that $\beta > \alpha$ in
(\ref{eq:input}). 
Let us consider a completely positive and trace-preserving (CPTP)
map\,\cite{Kraus:1983} 
\begin{equation}
\rho_{\rm in}
=
\proj{\psi}{\psi} 
\mapsto 
\rho_{\rm out}
=
 \sum_{\ell}
\Kop_{\ell}
\proj{\psi}{\psi} 
\Kop_{\ell}^{\dagger}
\label{eq:ltpcp}
\end{equation}
with a set of linear operator $\Kop^{\rm A}_{\ell}$ on $\mathbb{C}^{2}$,  
$\Kop_{\ell} = \Kop^{\rm A}_{\ell} \otimes \Iop_{d}$ 
and 
\(
\sum_{\ell}\Kop_{\ell}^{\dagger}\Kop_{\ell} 
= \Iop_{2}\otimes \Iop_{d}
\). 
The identity operator on $\mathbb{C}^{2}$ ($\mathbb{C}^{d}$) is denoted
by $\Iop_{2}$ ($\Iop_{d}$). 
The entanglement of $\rho_{\rm out}$ must be smaller than the
entanglement of $\rho_{\rm in}$ because this map is composed of only
local operators. 
However, a portion of $\rho_{\rm out}$, e.g.,  
\(
\Kop_{\ell}\proj{\psi}{\psi}\Kop_{\ell}^{\dagger}/
\Tr (\Kop_{\ell}\proj{\psi}{\psi}\Kop_{\ell}^{\dagger})
\) can be a more strongly entangled state than 
$\rho_{\rm in}$\,\cite{Brub;Leuchs:2007}.   
The local map (\ref{eq:ltpcp}) allows entanglement
distillation with probability  
\(
 \Tr (\Kop_{\ell}\proj{\psi}{\psi}\Kop_{\ell}^{\dagger})
\). 
Thus, the problem reduces to seeking a set of $\Kop_{\ell}^{\rm A}$'s
and how to apply them. 

Our goal can be thought of as entanglement distillation from a
single copy of a partially-entangled state. 
Bennett {\it et al.}\,\cite{Bennett;Schumacher:1996} suggested a
scheme to purify a maximally-entangled state from a partially-entangled
pure state using local partial-collapse
measurements\,\cite{Koashi;Ueda:1999}, i.e., 
\(
\Kop_{1}^{\rm A} = \sqrt{p}\proj{1}{1}
\)
and 
\(
\Kop_{2}^{\rm A} = \proj{0}{0} + \sqrt{1-p}\proj{1}{1}
\), 
with $0 \le p \le 1$. 
Kwiat {\it et al.}\,\cite{Kwiat;Gisin:2001} reported the experimental
realization of this scheme.  
Although one has to discard the quantum state when the outcome is
$\Kop_{1}^{{\rm A}\,\dagger}\Kop_{1}^{\rm A}$, one obtains a
maximally-entangled state when the outcome is 
$\Kop_{2}^{{\rm A}\,\dagger}\Kop_{2}^{\rm A}$.  
Several other entanglement distillation procedures have been proposed
(e.g., \cite{Maruyama;Nori:2008}) but using
many copies of a partially-entangled state. 
The demonstration of the methods with two copies of partially-entangled
states was reported in, e.g.,
\cite{Yamamoto;Imoto:2003,Pan;Zeilinger:2003}.  
The basic setting in this paper is the same as in
\cite{Bennett;Schumacher:1996,Kwiat;Gisin:2001} (i.e., the use 
of a single copy of a pure state and local measurements). 
Thus, hereafter, we will use the term ``distillation'' for describing
our protocol although we do not treat a large ensemble of mixed states
but just a single copy of a partially-entangled pure state. 

We propose a method for entanglement amplification with local weak
measurements and a single copy of a partially-entangled pure state. 
If the measurement is successful, we obtain a maximally-entangled state. 
We stress that our approach requires a small cost because a lot of copies
and two-body measurements are not needed. 
The type of the weak measurements used in our protocol is different from
the one in the proposal by Bennett {\it et al.} (so-called Procrutean
method)\,\cite{Bennett;Schumacher:1996}, and was demonstrated in, e.g, 
\cite{Iinuma;Hofmann:2011}. 
Although our approach is similar to the Procrutean method, the present
method does not completely destroy entanglement even if the measurement
fails. 
Thus, we can repeatedly apply weak measurements until we obtain a
maximally-entangled state. 
We also remark that when one has a physical system in which the
implementation of the weak measurements in
\cite{Bennett;Schumacher:1996} is difficult, one can use the present  
approach as an alternative method for entanglement distillation of a
single copy of a pure state.

\begin{figure}[htbp]
\centering
\caption{Schematic diagram of our approach to distill a
 maximally-entangled state from a partially-entangled initial state 
 $\ket{\psi}=\alpha\ket{0}\ket{\phi_{0}}+\beta\ket{1}\ket{\phi_{1}}$.
The measurements are performed only on system A. Each box indicates
 the measurement apparatus with the measurement strength
 $\epsilon_{n}$ ($n=1,\,2,\,\ldots,N$), see (\ref{eq:def_wmop}). 
When we obtain the outcome corresponding to
 $\Mop_{+}^{\dagger}\Mop_{+}$, we find a
 maximally-entangled state.}
\scalebox{0.32}[0.32]{\includegraphics{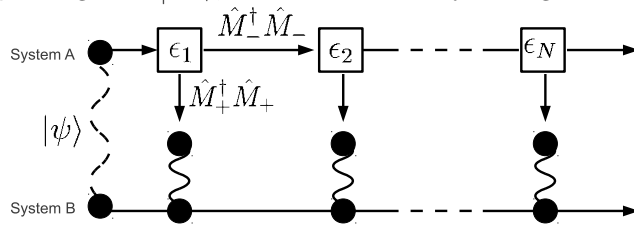}}
\label{fig:schematics} 
\end{figure}
\begin{figure}[htbp]
\centering
\caption{Application of our approach to
 $\alpha\ket{0}\ket{\phi_{0}}+\beta\ket{1}\ket{\phi_{1}}$ with 
 $\alpha^{2}=0.4$ and $\beta^{2}=0.6$. The horizontal axis is the number
 of measurements performed. The probability to obtain a 
 maximally-entangled state is shown in (a). The measurement
 strength is shown in (b). The sum of the net probabilities, which
 is defined as (\ref{eq:def_sum_netp}), is shown in (c). The dotted
 line indicates the total success probability, $2\alpha^{2}=0.8$. See 
 (\ref{eq:asymp_map}) and (\ref{eq:tot_s_prob}). A higher total
 success probability could be easily obtained by choosing $2\alpha^{2}$
 close to one.} 
\scalebox{0.65}[0.65]{\includegraphics{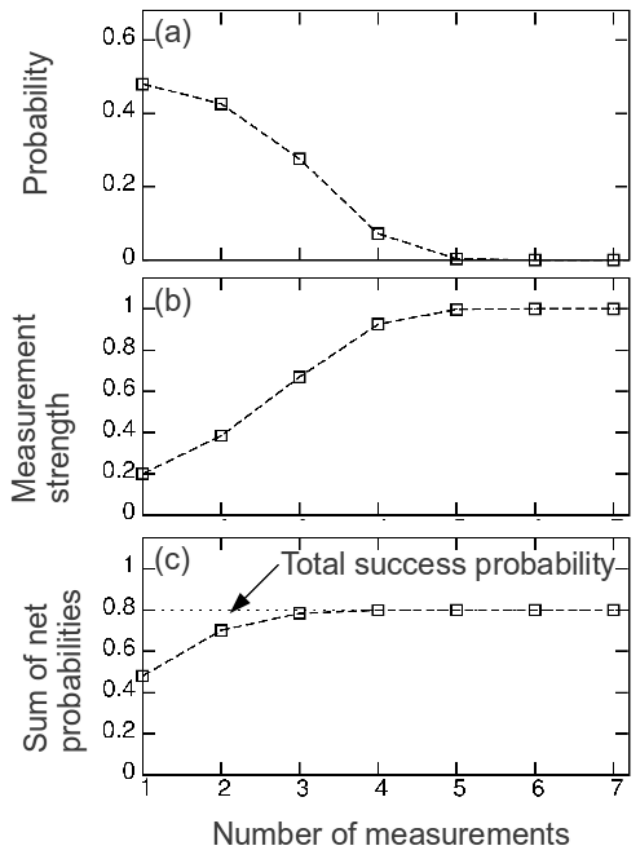}}
\label{fig:result_b0060} 
\end{figure}

\section{Entanglement amplification with local weak
 measurements} 
\label{sec:results}
Now we show a method to distill entanglement from a partially-entangled
pure initial state. 
The basic part of the scheme is the weak measurements
on system A given by 
\begin{eqnarray}
&&
 \Mop_{\pm }(\epsilon)
=
\sqrt{\frac{1 \pm \epsilon}{2}}
\proj{0}{0} 
+
\sqrt{\frac{1 \mp \epsilon}{2}}
\proj{1}{1}, 
\label{eq:def_wmop} \\
&&
 \Mop_{+}^{\dagger}\Mop_{+} + \Mop_{-}^{\dagger}\Mop_{-} 
= \Iop_{2},
\label{eq:id_wmop}
\end{eqnarray}
with measurement strength $\epsilon$ 
($0 \le \epsilon \le 1$). 
The set 
\(
\{\Mop_{\sigma}^{\dagger}\Mop_{\sigma}\}_{\sigma=+,-}
\) is a POVM\,\cite{Davies:1976,Kraus:1983,Peres:1993} on
$\mathbb{C}^{2}$.  
These measurements with tunable $\epsilon$ can be implemented in
various physical systems, as seen, e.g., in
\cite{Iinuma;Hofmann:2011,Nagali;Sciarrino:2011,Ota;Nori:2012}. 
We substitute $\Mop_{+}(\epsilon)$ and $\Mop_{-}(\epsilon)$,
respectively, into $\Kop^{\rm A}_{1}$ and $\Kop^{\rm A}_{2}$ in
(\ref{eq:ltpcp}). 
The resultant state is given by 
\(
 \rho_{\rm out}
=
p_{+} \proj{\psi_{+}}{\psi_{+}}
+
p_{-} \proj{\psi_{-}}{\psi_{-}}
\) 
with 
\(
\ket{\psi_{\pm}}
=
(\Mop_{\pm}\otimes \Iop_{d})\ket{\psi}
/ \sqrt{p_{\pm}}
\) and 
\(
p_{\pm}
=
\expec{\psi|
(\Mop_{\pm}^{\dagger}\Mop_{\pm}\otimes \Iop_{d})
|\psi}
\). 
The important point here is that $\ket{\psi_{+}}$ is a
maximally-entangled state when 
\(
\alpha^{2}(1+\epsilon)/2
=
\beta^{2}(1-\epsilon)/2
\). 
Assuming that $\beta > \alpha$, we find that the parameter
$\epsilon$ for obtaining a maximally-entangled state becomes 
\begin{equation}
\epsilon
=
\beta^{2}-\alpha^{2}
=
\sqrt{1 - S(\psi)}.
\label{eq:mes_cond}
\end{equation}
In contrast to a partial-collapse measurement\,\cite{Koashi;Ueda:1999},
the present measurement operators do not completely destroy the
entanglement of the input state even if the measurement fail. 
We also note that both the present approach and the Procrustean
method\,\cite{Bennett;Schumacher:1996} require a priori information of
the initial state (i.e., $\alpha$ and $\beta$). 

On the basis of the above arguments, we propose a probabilistic method
to make a maximally-entangled state from $\ket{\psi}$ with the successive
application of local weak measurements, as shown in
figure \ref{fig:schematics}. 
We prepare the $N$ weak measurement apparatus described by
(\ref{eq:def_wmop}) and (\ref{eq:id_wmop}). 
We stop the measurement process once we obtain the outcome
$\Mop_{+}^{\dagger}\Mop_{+}$.   
Otherwise, we move to the subsequent measurement. 
The input state in the $n$th weak measurement apparatus is 
\begin{eqnarray}
&&
\ket{\psi_{n}}
=
\frac{1}{p_{-}(\epsilon_{n-1})}
[\Mop_{-}(\epsilon_{n-1}) \otimes \Iop_{d}]\ket{\psi_{n-1}}, \\
&&
 p_{-}(\epsilon_{n-1})
=
 \expec{
\psi_{n-1} |
\Mop_{-}(\epsilon_{n-1})^{\dagger}
\Mop_{-}(\epsilon_{n-1}) \otimes \Iop_{d} |
\psi_{n-1}
},
\end{eqnarray}
if $n\ge 2$ and $\ket{\psi_{1}}=\ket{\psi}$, with $\beta>\alpha$. 
The expression of the weak measurements (\ref{eq:def_wmop}) allows the
form invariance of $\ket{\psi_{n}}$, i.e., 
\(
\ket{\psi_{n}}
=
\alpha_{n}\ket{0}\ket{\phi_{0}} 
+
\beta_{n}\ket{1}\ket{\phi_{1}} 
\) 
where $\alpha_{n}^{2}+\beta_{n}^{2}=1$ and $\alpha_{n},\beta_{n}>0$. 
In addition, we find that $\beta_{n} > \alpha_{n}$ if 
$\beta_{n-1} > \alpha_{n-1}$ and $\epsilon_{n}>0$. 
Then, when the measurement strength obeys the recurrence relation 
\begin{equation}
 \epsilon_{n} 
= \frac{2\epsilon_{n-1}}{1 + (\epsilon_{n-1})^{2}}
\quad
(n \ge 2),
\quad
\epsilon_{1} = \sqrt{1-S(\psi)},
\label{eq:mm_recurrence}
\end{equation}
the state corresponding to the outcome
\(
\Mop_{+}^{\dagger}(\epsilon_{n})\Mop_{+}(\epsilon_{n})
\otimes \Iop_{d}
\) is the maximally-entangled state 
\(
(\ket{0}\ket{\phi_{0}}+\ket{1}\ket{\phi_{1}})/\sqrt{2}
\). 
Indeed, the solution of (\ref{eq:mm_recurrence}) satisfies
(\ref{eq:mes_cond}), i.e.,  
\(
\epsilon_{n} = \sqrt{1-S(\psi_{n})}
\). 
The probability to obtain the outcome $\Mop_{+}^{\dagger}\Mop_{+}$ at the
$n$th measurement apparatus is 
\begin{equation}
  P_{n} 
= \frac{1}{2}[1 - (\epsilon_{n})^{2}].
\label{eq:prob_n}
\end{equation}
When $\beta < \alpha$, we can obtain the same result replacing the role
of $\Mop_{+}$ with $\Mop_{-}$. 
Since $\epsilon_{n}$ obeys the recurrence relation
(\ref{eq:mm_recurrence}), the information required is only $S(\psi)$. 

Figure \ref{fig:result_b0060} shows the probability $P_{n}$ and the
measurement strength $\epsilon_{n}$ for a specific input state.   
After a few steps, the probability decreases and approaches $0$, and the
measurement strength increases and approaches $1$.  
Actually, the recurrence relation (\ref{eq:mm_recurrence}) indicates that  
\(
\epsilon_{n} > \epsilon_{n-1}
\), and $\lim_{n \to \infty}\epsilon_{n}=1$. 
Therefore, the weak measurements asymptotically approach the von Neumann
measurements $\proj{0}{0}$ and $\proj{1}{1}$ for large $n$.  
Then, the state corresponding to the outcome  
\(
\Mop_{-}^{\dagger}\Mop_{-}\otimes \Iop_{d}
\) becomes $\ket{1}\ket{\phi_{1}}$. 

\begin{figure}[htbp]
\centering
\caption{Total success probability versus initial entanglement. The
 horizontal axis represents the linear entropy $S(\ket{\psi})$ of the
 state $\ket{\psi}$. 
The total success probability is defined as the
 asymptotic value of the sum of the net probability, as shown in
 figure \ref{fig:result_b0060}(c).}
\scalebox{0.58}[0.58]{\includegraphics{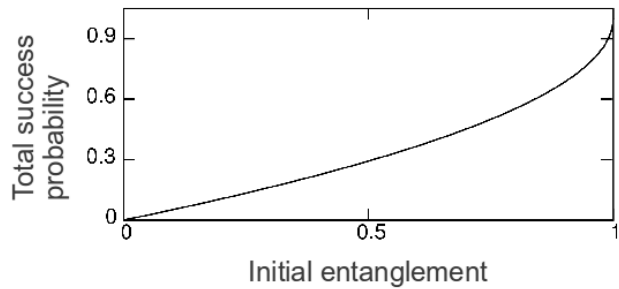}}
\label{fig:sp_coh} 
\end{figure}

Let us now characterize better the present approach.  
The $n$th measurement is used when the previous weak
measurement apparatus shows the outcome $\Mop_{-}^{\dagger}\Mop_{-}$. 
Thus, we may define the net probability to find a maximally
entangled state at the $n$th step as 
\begin{equation}
 P_{n}^{{\rm net}}
=
P_{n}\prod_{k=1}^{n-1}(1-P_{k}), 
\end{equation}
if $n \ge 2$ 
and 
\(
 P_{1}^{{\rm net}}
=
 P_{1}
\). 
We find that $P_{n-1}^{{\rm net}} > P_{n}^{{\rm net}}$. 
The sum of the net probabilities is a characteristic
quantity of this scheme. 
We define $P_{n}^{{\rm s}}$ as
\begin{equation}
 P_{n}^{{\rm s}} 
=
\sum_{m=1}^{n} P_{m}^{{\rm net}}. 
\label{eq:def_sum_netp}
\end{equation}
This quantity obeys the recurrence relation 
\(
P^{\rm s}_{n}
=
P^{\rm s}_{n-1} + (1-P_{n})P^{\rm s}_{n-1}
\) if $n \ge 2$ and 
\(
P^{\rm s}_{1} = P_{1}
\). 
Figure \ref{fig:result_b0060}(c) shows that $P^{\rm s}_{n}$ approaches a
specific value for large $n$.  
We confirm that this asymptotic value is $2\alpha^{2}$ by numerically
calculating $P^{\rm s}_{n}$ for the various initial states. 
We call this asymptotic value the total success probability of the
approach. 
Figure \ref{fig:sp_coh} shows that the total success probability is
large when $\alpha$ is close of $\frac{1}{2}$ [or $S(\psi)$ is close to
its maximum value]. 
Here, we show an alternative evaluation of the total success
probability. 
We consider an asymptotic map corresponding to the entire protocol.   
First, we focus on the fact that the entire protocol in the limit 
$n\to \infty$ can be regarded as a CPTP map 
\begin{equation}
 \proj{\psi}{\psi} 
\mapsto 
\rho_{\ast}
=
P_{\ast}\proj{\psi_{\rm MES}}{\psi_{\rm MES}}
+
(1 - P_{\ast})\proj{\chi}{\chi},
\label{eq:asymp_map}
\end{equation}
where 
\(
\ket{\psi_{\rm MES}}
=(\ket{0}\ket{\phi_{0}}+\ket{1}\ket{\phi_{1}})/\sqrt{2}
\), and 
\(
\ket{\chi} = \ket{1}\ket{\phi_{1}}
\). 
The positive coefficient $P_{\ast}$ corresponds to the total success
probability. 
This map must be a composite map of the weak measurements
(\ref{eq:def_wmop}). 
Therefore, we may write $\rho_{\ast}$ as 
\begin{eqnarray}
&&
 \rho_{\ast}
=
\sum_{\sigma=+,-}
(\Mop_{\ast,\sigma}\otimes \Iop_{d})
\proj{\psi}{\psi}
(\Mop_{\ast,\sigma}\otimes \Iop_{d})^{\dagger}, 
\label{eq:asymp_map_osr}
\\
&&
\Mop_{\ast,\pm}
=
 \sqrt{\frac{1  \pm \delta_{0}}{2}}
\proj{0}{0}
+
 \sqrt{\frac{1  \pm \delta_{1}}{2}}\proj{1}{1}, 
\label{eq:def_gwm} 
\end{eqnarray}
with 
\(
|\delta_{0}|,\,|\delta_{1}| \le 1
\). 
The measurement operators $\Mop_{\ast,\pm}$ correspond to generalized
weak (partial) measurements\,\cite{Paraoanu:2011:EPL,Paraoanu:2011:FP}.  
We find that (\ref{eq:asymp_map_osr}) represents a convex
combination between $\proj{\psi_{{\rm MES}}}{\psi_{{\rm MES}}}$ and 
$\proj{\chi}{\chi}$ if 
\begin{equation}
(\Mop_{\ast,+}\otimes \Iop_{d})\ket{\psi} 
\propto \ket{\psi_{{\rm MES}}}, 
\quad
(\Mop_{\ast,-}\otimes \Iop_{d})\ket{\psi} 
\propto \ket{\chi}.  
\end{equation}
Let us put
\(
\delta_{0}=1
\)
and
\(
\beta\sqrt{(1+\delta_{1})/2} = \alpha
\). 
Then, we find that 
\(
(\Mop_{\ast,+}\otimes \Iop_{d})\ket{\psi} 
=
\sqrt{2}\alpha \ket{\psi_{\rm MES}}
\). 
Thus, we show that 
\begin{equation}
 P_{\ast}= 2\alpha^{2}. 
\label{eq:tot_s_prob}
\end{equation}
In other words, we decomposed the map $(\ref{eq:asymp_map})$ with 
(\ref{eq:def_wmop}), as pointed out in \cite{Oreshkov;Brun:2005}. 
The measurement operators 
\(
\Mop_{\ast,+}
\) and 
\(
\Mop_{\ast,-}
\) for $\delta_{0}=1$ was experimentally demonstrated as a
partial-collapse measurement in
\cite{Kim;Kim:2009,Kwiat;Gisin:2001,Katz;Korotkov:2008}. 
We remark that the asymptotic map can be regarded as a single-shot partial
measurement and is equal to the Procrustean method in
\cite{Bennett;Schumacher:1996}. 
Actually, the total success probability $P_{\ast}$ is the same as the
yield of maximally-entangled states in
\,\cite{Bennett;Schumacher:1996}. 

Let us compare the present entanglement amplification protocol to the
Procrustean method in \cite{Bennett;Schumacher:1996} from the
viewpoint of physical implementation. 
The present approach requires the implementation of weak measurements that
are symmetric or balanced with respect to the measurement strength, as
seen in (\ref{eq:def_wmop}). 
Such symmetric weak measurements can be implemented in various physical
systems including linear optical and solid-state qubits, as shown, 
e.g., in \cite{Ota;Nori:2012}. 
In contrast to the symmetric measurements, the Procrustean
method uses non-symmetric ones. 
Namely, one uses (\ref{eq:def_gwm}) with $\delta_{0}=1$ and
$\delta_{1}\neq -1$. 
Although the non-symmetric partial measurements are naturally
realized in, e.g., Josephson phase qubits\,\cite{Katz;Korotkov:2008},
they are not always available in a general physical system. 
Thus, when one has a physical system in which the implementation of the
type of non-symmetric partial measurements that are used in
\cite{Bennett;Schumacher:1996} is difficult, one can use the present  
approach as an alternative method for entanglement distillation of a
single copy of a pure state. 

\section{Mixed states}
\label{sec:mixed}
We now extend our approach to a mixed initial state. 
Hereafter, we consider a two-qubit system (i.e., $d=2$). 
The study of a mixed initial state is motivated by experimental
considerations, as well as theoretical interest. 
One may desire to amplify entanglement of a quantum state in the
presence of decoherence, for example. 
We remark that one never distills a maximally-entangled state from a
single copy of a mixed state with local purification protocols, as shown
by Linden {\it et al.}\,\cite{Linden:Popescu:1998} and
Kent\,\cite{Kent:1998}. 
Hence, our goal in this section is to seek a protocol of increasing the
entanglement of an output state compared to a mixed initial state, not
purifying a maximally-entangled state. 
This section mainly focuses on a single-shot measurement protocol, not a
repeated application of weak measurements. 

The initial pure state (\ref{eq:input}) is replaced with the mixed state 
\begin{equation}
 \rho = \lambda \rho_{\rm s} + (1-\lambda)\proj{\psi}{\psi},
\label{eq:input_mix}
\end{equation}
where 
\(
 \ket{\psi} = \alpha\ket{00} + \beta\ket{11}
\) 
and 
\(
 0\le \lambda \le 1
\). 
This is a convex combination between a separable state $\rho_{\rm s}$
and a partially-entangled pure state $\ket{\psi}$. 
The coefficients $\alpha$ and $\beta$ are real and positive numbers in
the same way as (\ref{eq:input}). 
The weight of the separable part is quantified by $\lambda$. 
We note that an arbitrary density matrix on 
$\mathbb{C}^{2}\otimes \mathbb{C}^{2}$ can be written by the form
(\ref{eq:input_mix}), and this decomposition is uniquely
determined\,\cite{Lewenstein;Sanpera:1998}. 
We use the concurrence\,\cite{Wootters:1998,Audenaert;Moor:2001} to
evaluate the entanglement of a mixed state. 
Hereafter, the concurrence of $\rho$ is written as $C(\rho)$. 
We also use another quantity to estimate entanglement in addition to
the concurrence. 
When a mixed state is given as the form (\ref{eq:input_mix}), its
entanglement can be characterized by its pure-state part
$(1-\lambda)\proj{\psi}{\psi}$\,\cite{Lewenstein;Sanpera:1998}. 
Thus, an intuitive ``measure'' of entanglement is defined as 
\(
E(\rho) = (1-\lambda )S(\psi)
\), 
with the linear entropy (\ref{eq:l_entropy}). 
Using both $C(\rho)$ and $E(\rho)$, we examine the entanglement
amplification protocol for (\ref{eq:input_mix}). 
Hereafter, we assume that $\lambda \neq 0$ and $\beta > \alpha$.  

We use the same local two-outcome weak measurement in the case of a pure
initial state. 
As seen in section \ref{sec:results}, performing a single measurement
whose measurement operator is given by (\ref{eq:def_wmop}), we obtain 
\(
\rho_{\rm out} = R_{+}\rho_{+} + R_{-}\rho_{-}
\), with 
\(
R_{\pm}
=
\lambda w_{{\rm s},\pm} 
+ 
(1-\lambda) p_{\pm}
\), 
\(
\rho_{\pm}
=
(\Mop_{\pm}\otimes \Iop_{2})
\rho
(\Mop_{\pm}^{\dagger}\otimes \Iop_{2})
/R_{\pm}
\), 
and 
\(
w_{{\rm s},\pm}
=
\tr[(\Mop_{\pm}^{\dagger}\Mop_{\pm}\otimes \Iop_{2})
\rho_{\rm s}]
\). 
Our aim is to increase the entanglement of $\rho_{+}$, compared with
$\rho$. 
This density matrix has the same form as (\ref{eq:input_mix}),  
\begin{equation}
 \rho_{+} 
= 
\lambda_{+} 
\frac{
(\Mop_{+}\otimes \Iop_{2})
\rho_{\rm s} 
(\Mop_{+}^{\dagger}\otimes \Iop_{2}) }
{w_{{\rm s},+}}
+
(1-\lambda_{+})
\proj{\psi_{+}}{\psi_{+}},
\label{eq:out_p_mix} 
\end{equation}
with 
\begin{equation}
 \lambda_{+} = \frac{w_{{\rm s},+}}{R_{+}}\lambda.
\end{equation}
We examine $E(\rho)$ and $E(\rho_{+})$ to obtain a simple criterion for
the entanglement amplification. 
Since $\rho_{\rm s}$ is a separable density matrix and the weak
measurement is local, the first term in (\ref{eq:out_p_mix}) is the
separable part of $\rho_{+}$. 
Therefore, $E(\rho_{+})=(1-\lambda_{+})S(\psi_{+})$. 
As shown in section \ref{sec:results}, $\ket{\psi_{+}}$ is a
maximally-entangled state when the measurement strength satisfies
(\ref{eq:mes_cond}). 
It means that $S(\psi_{+})>S(\psi)$. 
To ensure $E(\psi_{+}) \ge E(\psi)$, we require 
$\lambda_{+} \le \lambda$. 
This relation leads to 
\begin{equation}
A_{{\rm s},z} \le - \sqrt{1-S(\psi)} = \alpha^{2}-\beta^{2},
\label{eq:ano_cond_mix}
\end{equation}
where we have used (\ref{eq:mes_cond}). 
We have defined $A_{{\rm s},z}$ as 
\(
A_{{\rm s},z} 
= 
\tr[
(\sigma_{z}\otimes \Iop_{2})\rho_{\rm s}
]
\), 
where 
\(
\sigma_{z} = \proj{0}{0} - \proj{1}{1}
\). 
Therefore, we expect that the method for a pure initial state works
even for a mixed initial state if this inequality is satisfied. 
We will check this simple criterion for specific examples using
calculations of the concurrence. 
Then, our numerical calculations of the concurrence will show that
(\ref{eq:ano_cond_mix}) is the sufficient condition for 
$C(\rho_{+}) \ge C(\rho)$.  

\subsection{Example 1: Pure dephasing}
Let us consider the case when one of the subsystems (e.g., system A)
travels via a noisy channel after the preparation of $\ket{\psi}$, for
example. 
Assuming that the system undergoes pure dephasing, the corresponding
CPTP map is written as 
\(
\rho
\mapsto
(1-u) \rho 
+ u 
(\sigma_{z}\otimes \Iop_{2})\rho (\sigma_{z}\otimes \Iop_{2})
\), 
with $0\le u \le 1/2$. 
Thus, the initial mixed state becomes 
\begin{equation}
 \rho^{({\rm pd})}
=
\alpha^{2}\proj{00}{00}
+ \mu \proj{00}{11} 
+ \mu \proj{11}{00}
+ \beta^{2} \proj{11}{11},
\label{eq:def_pd}
\end{equation}
with $\mu = \alpha\beta (1-2u)$. 
We can rewrite (\ref{eq:def_pd}) as 
\(
\rho^{({\rm pd})}
=
[1-(\mu/\alpha\beta)]
(
\alpha^{2}\proj{00}{00}
+
\beta^{2}\proj{11}{11}
)
+
(\mu/\alpha\beta)\proj{\psi}{\psi}
\). 
Non-classical features of this sort of density matrices are
characterized well by the off-diagonal element
$\mu$\,\cite{Ashhab;Nori:2007}. 
In fact, the concurrence is 
\(
C(\rho^{({\rm pd})}) = 2\mu
\). 
Let us check the inequality (\ref{eq:ano_cond_mix}). 
Since 
\(
A_{{\rm s},z} = \alpha^{2} - \beta^{2}
\) 
and 
\(
\sqrt{1-S(\psi)}=\beta^{2}-\alpha^{2}
\), 
the left hand side of (\ref{eq:ano_cond_mix}) is zero and the inequality
is satisfied. 
Thus, we expect that the approach for a pure initial state leads to an
amplification of the entanglement of $\rho^{({\rm pd})}$. 
We calculate the concurrence to confirm that this prediction is correct. 
The concurrence of 
\(
\rho^{({\rm pd})}_{+}
\propto 
(\Mop_{+}\otimes \Iop_{2})
\rho^{({\rm pd})}
(\Mop_{+}^{\dagger}\otimes \Iop_{2})
\) is 
\begin{equation}
 C(\rho_{{\rm pd},+}) 
=
\frac{\mu}{\alpha\beta}
>
2\mu
=
C(\rho^{({\rm pd})}).
\end{equation}
Therefore, entanglement amplification is achieved. 
Furthermore, we can find that the repeated application still works. 
The input density matrix for the $n$th measurement apparatus is the
state corresponding to the outcome 
\(
\Mop_{-}^{\dagger}(\epsilon_{n-1})\Mop_{-}(\epsilon_{n-1})
\). 
We denote this state as $\rho_{n}$. 
We note that $\rho_{1}=\rho^{({\rm pd})}$. 
When we obtain the outcome 
\(
\Mop_{+}^{\dagger}(\epsilon_{n})\Mop_{+}(\epsilon_{n})
\), the measurement is considered to be successful. 
We write the resultant state as $\rho^{({\rm s})}_{n}$. 
The measurement strength obeys the recurrence relation
(\ref{eq:mm_recurrence}) with $\epsilon_{1}=\sqrt{1-S(\psi)}$. 
The application to $\rho$ is straightforward because $\rho_{n}$ and
$\rho^{({\rm s})}_{n}$ have the same 
form as $\rho$ (i.e., only the coefficients change). 
We can show that 
\(
C(\rho^{({\rm s})}_{n}) > C(\rho_{n})
\)
and 
\(
C(\rho_{n}) < C(\rho_{n-1})
\). 
Eventually, we find that 
\(
C(\rho^{({\rm s})}_{n}) = C(\rho^{({\rm s})}_{n-1})
\). 
Therefore, at every value for the outcome 
$\Mop_{+}^{\dagger}\Mop_{+}$, we obtain a mixed state with the
concurrence  
\(
C(\rho^{({\rm s})}_{1}) 
=
\mu/\alpha\beta 
\). 
Remarkably, we find that 
\(
\mu/\alpha\beta > C(\rho^{({\rm pd})})
\) because 
\(
2\alpha\beta < 1
\). 
The resultant entanglement is bounded by $\mu/\alpha\beta$, which is
regarded as a measure of the purity of $\rho^{({\rm pd})}$. 

\begin{figure}[htbp]
\centering
\caption{(Color online) Characteristics of the entanglement
 amplification method for the mixed state associated with amplitude
 damping, (\ref{eq:def_ad}). The horizontal axis is the 
 linear entropy of the reduced density matrix of the pure-state part,
 $S(\tilde{\psi})$. The vertical axis is the weight of the separable part,
 $(1-u)\beta^{2}$. (a) Concurrence of the input state $\rho^{({\rm ad})}$
 (left panel) and the output state $\rho^{({\rm ad})}_{+}$ (right
 panel). Using (\ref{eq:mes_cond}), the measurement strength is given by 
 $\sqrt{1-S(\tilde{\psi})}$. (b) Profile of 
${\rm sgn}[C(\rho^{({\rm ad})}_{+})-C(\rho^{({\rm ad})})]$. 
When this quantity is positive, the entanglement amplification is
 achieved (red-colored region). Otherwise, the protocol fails
 (blue-colored region). }
\scalebox{0.83}[0.83]{\includegraphics{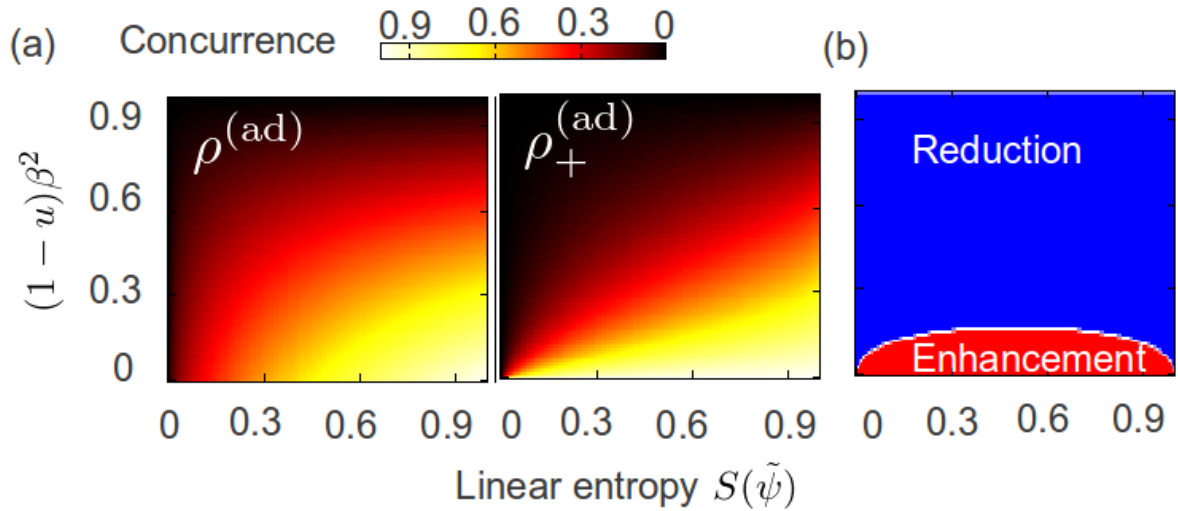}}
\label{fig:ampd} 
\end{figure}

\subsection{Example 2: Amplitude damping}
The second example is associated with amplitude-damping. 
If this effect occurs on system A, the corresponding CPTP map is written
as 
\(
\rho
\mapsto
\sum_{\ell=1,2}
(\Eop_{\ell}\otimes \Iop_{2})
\rho
(\Eop_{\ell}^{\dagger}\otimes \Iop_{2})
\), 
with 
\(
\Eop_{1}
=
\proj{0}{0} + \sqrt{u}\proj{1}{1}
\), 
\(
\Eop_{2}
=\sqrt{1-u} \proj{0}{1}
\), 
and 
\(
0 \le u \le 1
\). 
Thus, the mixed state becomes
\begin{equation}
 \rho^{({\rm ad})}
=
(1-u)\beta^{2}\proj{01}{01}
+
[1-(1-u)\beta^{2}]
\proj{\tilde{\psi}}{\tilde{\psi}},
\label{eq:def_ad}
\end{equation}
with 
\(
\ket{\tilde{\psi}}
=
(\alpha\ket{00} + \sqrt{u}\beta\ket{11})
/
\sqrt{1-(1-u)\beta^{2}}
\). 
Assuming that $\sqrt{u}\beta > \alpha$, we check the inequality
(\ref{eq:ano_cond_mix}). 
Since 
\(
\sqrt{1-S(\tilde{\psi})}
=
(u\beta^{2}-\alpha^{2})/(u\beta^{2}+\alpha^{2})>0
\) 
and 
\(
A_{{\rm s},z}
=1
\), 
(\ref{eq:ano_cond_mix}) is not satisfied. 
Thus, we expect that the method for a pure initial state does not work. 
A numerical calculation of the concurrence gives more quantitative
characterization of our protocol, as seen in figure \ref{fig:ampd}. 
Varying the weight of the separable part $(1-u)\beta^{2}$ and the linear
entropy of the reduced density matrix of the pure-state part
$S(\tilde{\psi})$, the concurrences of $\rho^{({\rm ad})}$ [left panel of
figure \ref{fig:ampd}(a)] and 
\(
\rho^{({\rm ad})}_{+} 
\propto 
(\Mop_{+}\otimes \Iop_{2})
\rho^{({\rm ad})}
(\Mop_{+}^{\dagger}\otimes \Iop_{2})
\) [right panel of figure \ref{fig:ampd}(a)] are calculated. 
Furthermore, the sign of their difference (
\(
{\rm sgn}[
C(\rho^{({\rm ad})}_{+})
-
C(\rho^{({\rm ad})})
]
\)
) is evaluated, as seen in figure \ref{fig:ampd}(b). 
Figure \ref{fig:ampd}(b) shows that the method for a pure state leads to
the reduction of the entanglement in wide parameter region. 
Nevertheless, when the purity of $\rho^{({\rm ad})}$ is large, the
entanglement is amplified. 

\begin{figure}[htbp]
\centering
\caption{(Color online) Characteristics of the entanglement
 amplification method for the initial state (\ref{eq:rd}). 
The horizontal axis is the linear entropy of the reduced density matrix
 of a pure-state part, $S(\psi)$. The vertical axis is the weight of the
 separable part, $\lambda$. (a) Concurrence of the input state
 $\rho^{({\rm rnd})}$ (left panel) and the output state 
$\rho^{({\rm rnd})}_{+}$ (right panel). 
Using (\ref{eq:mes_cond}), the measurement strength is given by 
 $\sqrt{1-S(\psi)}$. (b) Profile of 
${\rm sgn}[C(\rho^{({\rm rnd})}_{+})-C(\rho^{({\rm rnd})})]$. When this
 quantity is positive, the entanglement amplification is achieved
 (red-colored region). Otherwise, the protocol fails (blue-colored
 region). In the white-colored region, the concurrence does not change
 because the initial mixed state (\ref{eq:rd}) is separable.} 
\scalebox{0.83}[0.83]{\includegraphics{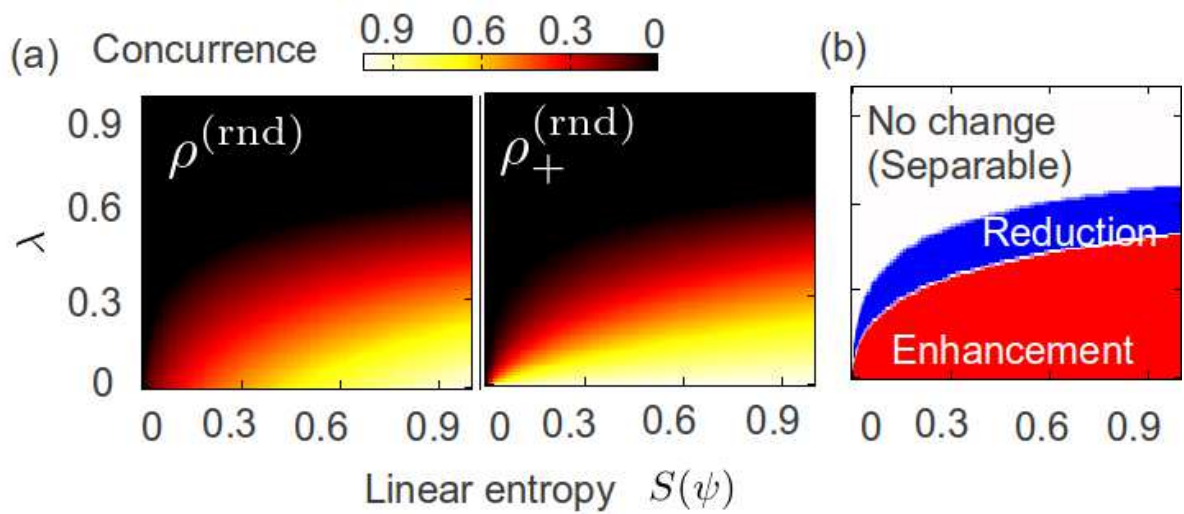}}
\label{fig:random} 
\end{figure}

\subsection{Example 3: Maximally-mixed state}
The parameter region in which the entanglement amplification for mixed
states works depends on the value of $A_{{\rm s},z}$. 
Let us examine the case when the separable state in (\ref{eq:input_mix})
is the maximally-mixed state, for example. 
The initial mixed state becomes
\begin{equation}
 \rho^{({\rm rnd})}
=
 \lambda \frac{1}{4}\Iop_{2}\otimes \Iop_{2}
+
 (1-\lambda)\proj{\psi}{\psi}. 
\label{eq:rd}
\end{equation}
We find that $A_{{\rm s},z}=0$. 
Therefore, the inequality (\ref{eq:ano_cond_mix}) is not satisfied
except for $S(\psi)=1$. 
In order to evaluate quantitative behaviors of the protocol, let us
examine the concurrence. 
Figure \ref{fig:random}(a) shows the concurrences of $\rho^{({\rm rnd})}$
(left panel) and 
\(
\rho^{({\rm rnd})}_{+}
\propto
(\Mop_{+}\otimes \Iop_{2})
\rho_{{\rm rd}}
(\Mop_{+}^{\dagger}\otimes \Iop_{2})
\) (right panel). 
The sign of their difference 
(\(
{\rm sgn}[
C(\rho^{({\rm rnd})}_{+})
-
C(\rho^{({\rm rnd})})]
\)
) is shown in figure \ref{fig:random}(b). 
Again, we find that the method for a pure state fails in wide
parameter region. 
However, compared with figure \ref{fig:ampd}(b), the entanglement
amplification is achieved in much wider region. 
Thus, smaller values of $A_{{\rm s},z}$ lead to
higher success of the protocol for various parameters. 
We also find that from (\ref{eq:ano_cond_mix}) negative values of
$A_{{\rm s},z}$ are preferable for the entanglement amplification. 

\begin{figure}[htbp]
\centering
\caption{(Color online) Profiles of concurrence differences between the
 output and the initial states, varying $A_{{\rm s},z}$. Each point is a
 mean value over $10000$ separable states. The horizontal axis is the
 linear entropy of the reduced density matrix of the pure-state part. 
The vertical axis is the weight of the separable part. See
 (\ref{eq:input_mix}). When the differences are positive (red-colored
 regions), the entanglement amplification is achieved. Otherwise, the
 protocol fails (blue-colored regions). 
In the white-colored regions, the concurrence does not change because
 the mixed initial state is separable. 
In the right regions of the the vertical (green) lines, 
the inequality (\ref{eq:ano_cond_mix}) is satisfied.}
\scalebox{0.8}[0.8]{\includegraphics{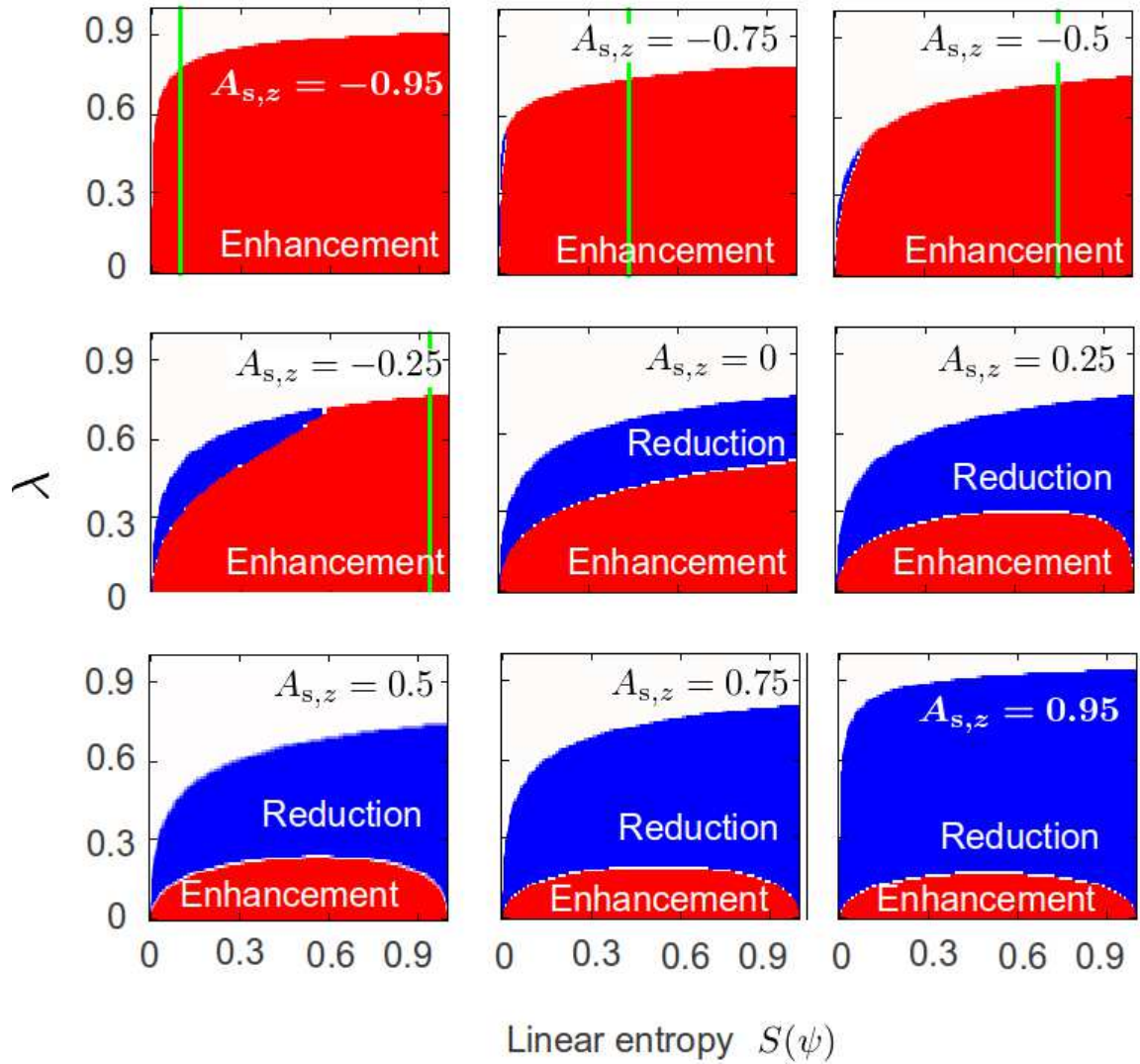}}
\label{fig:pdiagram} 
\end{figure}

\subsection{Systematic examinations}
For a more systematic analysis for mixed initial states, we randomly
generate separable density matrices. 
Thus, we examine entanglement amplification for various density matrices
given by (\ref{eq:input_mix}), with fixed $\ket{\psi}$. 
A separable state on $\mathbb{C}^{2}\otimes \mathbb{C}^{2}$ is
parametrized by $15$ real parameters. 
Using the Pauli matrices, we can express it as
\begin{equation}
 \rho_{\rm s}
=
\frac{1}{4}
\left(
\Iop_{2}\otimes \Iop_{2} 
+
\sum_{i}
A_{{\rm s},i}\sigma_{i}\otimes \Iop_{2}
+
\sum_{j}
B_{{\rm s},j}\Iop_{2}\otimes \sigma_{j}
+
\sum_{i,j}
C_{{\rm s},ij}
\sigma_{i}\otimes \sigma_{j}
\right),
\label{eq:sep_s}
\end{equation}
with $i,j=x,y,z$ and 
\(
A_{{\rm s},i}, B_{{\rm s},j},C_{{\rm s},ij}\in\mathbb{R}
\). 
The positivity and the separability of (\ref{eq:sep_s}) give constraints
with respect to the real parameters 
$A_{{\rm s},i}$, $B_{{\rm s},j}$, and $C_{{\rm s},ij}\in\mathbb{R}$. 
The former is straightforwardly checked using 
Newton's formula (see, e.g., \cite{Kimura:2003}). 
The latter is checked by the positive partial transpose
criterion\,\cite{Peres:1996,Horodecki;Horodecki:1996}. 
We calculate the mean values of the concurrences of $\rho$ and
$\rho_{+}$ with respect to randomly-generated separable states. 
Then, we evaluate 
\(
{\rm sgn}[\overline{C}(\rho_{+})-\overline{C}(\rho)]
\), 
where $\overline{C}$ represents the mean value of the concurrence. 
To examine the $A_{{\rm s},z}$ dependence of the protocol, we randomly
generate $\rho_{{\rm s}}$ fixing $A_{\rm s}$ as a specific value. 
We generate $10000$ samples for each value of $A_{{\rm s},z}$ using the
Mersenne Twister method\,\cite{Matsumoto;Nishimura:1998}. 
Figure \ref{fig:pdiagram} shows \(
{\rm sgn}[\overline{C}(\rho_{+})-\overline{C}(\rho)]
\), varying $A_{{\rm s},z}$. 
When this difference is positive (red-colored regions), the entanglement
amplification is achieved. 
We note that in the white-colored region the entanglement is not changed
because the mixed initial states are separable. 
The vertical (green) lines 
(for $A_{{\rm s},z}=-0.95,\,-0.75,\,-0.55,\,-0.35$) corresponds to the
equality of (\ref{eq:ano_cond_mix}),  
\(
S(\psi) = 1 -(A_{{\rm s},z})^{2}
\). 
In the right regions of the lines, the inequality
(\ref{eq:ano_cond_mix}) is satisfied. 
Thus, figure \ref{fig:pdiagram} shows that (\ref{eq:ano_cond_mix}) is
the sufficient condition for $C(\rho_{+})\ge C(\rho)$. 
Furthermore, when $A_{{\rm s},z}$ is a large negative value, the
entanglement is enhanced in wide parameter region. 

Summarizing the above arguments, the inequality (\ref{eq:ano_cond_mix})
is the sufficient condition for the entanglement amplification. 
In addition, the method for a pure intial state can still lead to
entanglement amplification even for a mixed initial state if the
separable part of the mixed state fulfills $A_{{\rm s},z}<0$.  
The reason why negative values of $A_{{\rm s},z}$ are preferable for the
entanglement amplification is understood by considering the role of the
measurement operator $\Mop_{+}$ in the present protocol. 
When the measurement strength fulfills (\ref{eq:mes_cond}), $\Mop_{+}$
reduces components with respect to $\ket{1}$ of an input state $\rho$
(e.g., $\expec{11 |\rho | 11}$). 
A large negative value of $A_{{\rm s},z}$ implies that  
\(
\tr[(\proj{1}{1}\otimes \Iop_{2})\rho_{\rm s}]
\) is a predominant component of the separable part of the input state. 
Therefore, the measurement operator $\Mop_{+}$ considerably decreases
the weight of the separable part in the input density matrix. 
We remark that positive values of $A_{{{\rm s},z}}$ are preferable for
the entanglement amplification when $\alpha > \beta$ and the measurement
strength satisfies $\epsilon=\alpha^{2}-\beta^{2}(>0)$. 

\section{Implementing the protocol with general measurements}
\label{sec:discussion}
We mention that no specific type of weak measurements is essential for
this approach. 
We can use a generalized two-outcome measurement, 
\begin{eqnarray}
&&
\Mop_{1} = \sqrt{p}\proj{0}{0} + \sqrt{q}\proj{1}{1}, \\
&&
\Mop_{2} = \sqrt{1-p}\proj{0}{0} + \sqrt{1-q}\proj{1}{1}, 
\end{eqnarray}
with $0\le p \le 1$ and $0 \le q \le 1$. 
We note that in the Procrustean method one of the two parameters is
fixed as $1$. 
The two real parameters $p$ and $q$ may be determined by the
requirements $\ket{\psi}=\ket{\psi_{\rm MSE}}$ and $\lambda_{+} \le
\lambda$, given as, respectively, 
\begin{eqnarray}
&&
p\alpha^{2} = q\beta^{2}, \\ 
&&
(p-q)(\sqrt{1-S(\psi)} + A_{{\rm s},z}) \le 0.
\end{eqnarray}
For pure initial states, the latter condition is not necessary. 
Adjusting the two real parameters depending on the initial
partially-entangled state and the separable part, we may achieve
entanglement amplification with some probability. 
Alternatively, we may use a continuous weak measurement. 
Continuously monitoring system A with very weak measurement strength,
the entanglement of formation of the total system may fluctuate and take
its maximal value $1$ at some point in time. 
We note that both of them have the same efficiency as the Procrustean
method. 

\section{Summary}
\label{sec:summary}
We have proposed a local probabilistic protocol to make a maximally-entangled
state from a partially-entangled pure state by applying elementary
two-outcome weak measurements. 
The present approach requires only the implementation of local
measurements and the information of the initial entanglement. 
The probability to find the maximally-entangled state at each step
decreases to zero with increasing measurement number. 
In addition, we found that the present approach corresponds to a
decomposition of a two-outcome POVM which generates maximally-entangled
or separable states as the measurement results.  
The success probability of the map corresponding to this POVM is
high if the initial entanglement is large. 
The case for a mixed initial state was also examined. 
We found that the method for a pure state still leads to entanglement
amplification, but not purifying a maximally-entangled state, if the
separable part satisfies a certain condition. 
Several future issues will contribute to the development of weak
measurement-based quantum control and the illustration of interesting
and useful properties of POVM's. 

An interesting application example of this method is a quantum
network system in which system A is a mediator
(e.g., photon) and system B is a quantum memory. 
In this case, it is not easy to perform controlled operations between
the two systems after their initial preparation. 
A local-operation-based approach will be effective for such a situation. 

\ack
We thank P. D. Nation, J. R. Johansson, N. Lambert, T. Ichikawa,
 K. Yuasa, and S. Pascazio for their very useful comments.  
YO is partially supported by the Special Postdoctoral Researchers Program,
RIKEN. 
The calculations were partially performed by using the RIKEN Integrated
Cluster of Clusters facility. 
SA and FN acknowledge partial support from LPS, NSA, ARO, NSF
grant No.~0726909, JSPS-RFBR contract number 09-02-92114, Grant-in-Aid
for Scientific Research (S), MEXT Kakenhi on Quantum Cybernetics and
JSPS via its FiRST program.

\end{document}